\documentclass[12pt,a4paper]{article}
\usepackage{amsmath}
\usepackage{amssymb}
\begin{document}
 \begin{titlepage}
  \begin{flushright}
IFUP--TH/2010-38-r\\
  \end{flushright}
%~
%\vskip .8truecm
 \begin{center}
\Large\bf
Riemann-Hilbert treatment of Liouville theory on the torus
\end{center}
\vskip 1.2truecm
\begin{center}
{Pietro Menotti} \\
{\small\it Dipartimento di Fisica, Universit{\`a} di Pisa and}\\
{\small\it INFN, Sezione di Pisa, Largo B. Pontecorvo 3, I-56127}\\
{\small\it e-mail: menotti@df.unipi.it}\\
 \end{center}
\vskip 1.2truecm
\centerline{October 2010}
                
\vskip 1.2truecm
                                                              
\begin{abstract}
We apply a perturbative technique to study classical Liouville theory on the
torus. After mapping the problem on the cut-plane we give the perturbative
treatment for a weak source. 
When the torus reduces to the square the problem is
exactly soluble by means of a quadratic transformation in terms of
hypergeometric functions. We give general formulas for the deformation of a
torus and apply them to the case of the deformation of the square.  One can
compute the Heun parameter to first order and express the solution in
terms of quadratures.  In addition we give in terms of quadratures of
hypergeometric functions the exact symmetric Green function on the square on 
the background generated by a one point source of arbitrary strength.

\end{abstract}

\end{titlepage}

\eject

\section{Introduction}

Liouville theory has been widely studied both at the classical level (see
e.g. \cite{picard,poincare,lichtenstein,troyanov,bilal}) 
and at the quantum level
\cite{bilal,ZZsphere,ZZpseudosphere,ZZdisk,dorn,teschner,nakayama}.  
Most of the attention was devoted to the sphere
and disk topologies \cite{ZZsphere,ZZpseudosphere,ZZdisk,dorn,teschner}.
Recently some notable progress have been performed for the torus topology
\cite{onofri,litvinov,hadasz,gaberdiel1,poghossian}.

Important results regarding the four point conformal correlation functions 
on the sphere  and their relation to the one point
function on the torus have been obtained 
\cite{onofri}.  
Results related to conformal
quantum field theory on the tours and to vortex theory on the torus are found 
in \cite{alba,gaberdiel} and \cite{olesen,akerblom}. 

In \cite{MV,MT,menotti} the sphere and the disk topologies were studied 
within the
standard functional formulation of quantum field theory, i.e. by first
computing a classical background and then integrating on the fluctuations
about it.  The three point case on the sphere, the one point case on the disk
and also the two point case on the disk when one source is weak were studied
in \cite{MV,MT,menotti}.

In this way it was possible to confirm the first few terms given by
the bootstrap approach \cite{ZZsphere,ZZpseudosphere,ZZdisk,teschner} 
and also to compute some higher-point correlation functions when one source 
is weak \cite{MT}. 
The classical four point case on the sphere was studied in \cite{MV} 
when one of the sources is weak, in which case the problem is soluble 
by quadratures and
as a byproduct one obtains the exact Green function on the background of three
sources of arbitrary strength, provided they satisfy Picard's inequalities.

In this paper we shall consider Liouville theory at the classical level on the
torus topology first in the general case and then we shall specialize to the
one point source at the classical level. Even this simple configuration is not
trivial. The case of the sphere, due to Picard inequalities, is meaningful
only when the sources are at least three and the three point case is soluble
in terms of hypergeometric functions through a well known procedure
\cite{bilal,MV}. The
simplicity of the solution of the three point case on the sphere is due to the
fact that the accessory parameters are completely fixed by the Fuchs
conditions and on the sphere the monodromy matrices are completely
known \cite{batemanII,bilal,MV} in terms of gamma-functions.

On the torus through a transformation involving the Weierstrass 
${\wp}$-function it is possible \cite{WW,batemanII,onofri} 
to map the problem on the cut-plane but already with
one point source the Fuchsian differential equation has four singularities and
thus it becomes a special case of the Heun equation. A large literature exists
on the Heun equation (see e.g.  \cite{erdelyi,maier,slavyanov,hejhal}).  
The main problem in dealing with the $1$-point problem on the torus is the
determination of the Heun parameter (accessory parameter).  Here the Fuchs
conditions are not sufficient and the Heun parameter, which we shall call
$\beta$ has to be determined along with an other parameter $k$ which also
enters the $3$-point solution on the sphere, by imposing the monodromic nature
of the solution. This is a variant of the XXI Hilbert problem
\cite{bolibrukh,CMS}.
A further shortcoming is the lack 
of analytical expressions for the monodromy matrices of the solutions 
of the Heun equation \cite{erdelyi,hejhal,hempel}.

Obviously the exact solution cannot be given in general in terms of familiar
functions like the hypergeometric functions, but the knowledge of $\beta$ from
the data of the problem would be a decisive step. After $\beta$ is known, the
parameter $k$ is easily computed.

One exception is the square configuration where by symmetry reasons the
accessory parameter vanishes. Then by a quadratic transformation the exact
solution is expressible in terms of hypergeometric functions. One can apply a
perturbative technique developed in \cite{MV} to work out the
solution on a deformed square and on the square in which a symmetric
distribution of weak sources is added.

The structure of the paper is the following: in Sec.\ref{general} we give the
general setting of the problem. In Sec.\ref{periods} we report general
relations between the periods and the root $e_k$ of a third order
equation which will be useful in the following. 
In Sec.\ref{pertonepoint} we give the perturbative treatment of
the one point function. A byproduct
is a derivation of the Green function on the torus which is alternative to the
standard one \cite{itzykson,difrancesco,weil}. 
In Sec.\ref{degreesoffreedom} we
show at the non perturbative level how the degrees of freedom appearing in the
differential equation are sufficient to realize the monodromic
solution. In Sec.\ref{square} we give the treatment of the
square in terms of hypergeometric functions. 
Secs.\ref{generaldeformation},\ref{squaredeformation} deal 
with the
perturbation induced by a general deformation of the torus and its application
to the deformation of the square. Finally in Sec.\ref{backgroundgreen} we
compute in terms of quadratures the exact symmetric Green function on the 
square in the background generated by an arbitrary charge.

Here we bound ourselves the the classical aspects;
the technique of the expansion of quantum Liouville theory around classical
solutions (semiclassical expansion) was applied with success to the case of
the sphere topology \cite{MV} and to the disk topology
\cite{MT,menotti} for computing various correlation functions.

\section{General setting}\label{general}

We write Liouville equation on the torus with $N$ point sources in the form 
\begin{equation}\label{philiouville} 
-\partial_z\partial_{\bar
z}\phi+e^\phi= 2\pi \sum_n \eta_n\delta^2(z-z_n). 
\end{equation}
It will be useful for the following to define $\lambda_n = 1 -2 \eta_n$. The
$\eta_n$ are subject to the Picard's inequalities
$$
\eta_n<\frac{1}{2},~~~~{\rm or}~~~~0<\lambda_n
$$
and
$$
\sum_n \eta_n >0.
$$
The first inequality is requested by the local finiteness of the area  
while the second is due to the Gauss-Bonnet theorem.

A direct transcription of the problem (\ref{philiouville}) into a linear 
differential equation obtained by developing the torus on the plane would 
generate an equation of the type
$$
y'' + Q(z) y=0
$$
with $Q(z)$ containing an infinite number of poles due to the infinite images
of the $n$ sources and this is not a Fuchsian problem.

One performs the change of variables \cite{batemanII,WW}
$u=\wp(z)$, being $\wp$ the
Weierstrass function related to the torus whose periods we shall denote
by $2\omega_1$, $2\omega_2$. The above transformation due to $\wp(z)=\wp
(-z)$ gives a two valued map from the cut-plane $u$ to
the fundamental region in the $z$-plane which describes the torus.  The
original Liouville equation is transformed into
\begin{equation}\label{varphiliouville} 
-\partial_u\partial_{\bar u}\varphi + e^{\varphi} = 2\pi
\sum_n \eta_n\delta^2(u-u_n) + 
       \frac{\pi}{2}\left(\delta^2(u-e_1)+\delta^2(u-e_2)+\delta^2(u-e_3)
\right)
\end{equation}
where 
\begin{equation}\label{phivarphi}
\varphi = \phi+\log (\frac{dz}{du}\frac{d\bar z}{d\bar u}).
\end{equation}
The points $e_1,e_2,e_3$ are the images in the $u$-plane of the points
$\omega_1$, $\omega_2$ and $\omega_3\equiv \omega_1+\omega_2$ and are
subject to the condition $e_1+e_2+e_3=0$.
If one source in eq.(\ref{philiouville}) is placed at the origin, say $z_0=0$
such  a source does not appear in eq.(\ref{varphiliouville}) because the
transformation takes it to infinity and it is taken into
account by the behavior of $\varphi$ at infinity as dealt with in 
eq.(\ref{beequation}) below. 
The new sources appearing at $e_1,e_2,e_3$ are due to the Laplacian of
eq.(\ref{varphiliouville}) acting on
the logarithms appearing in the r.h.s. of eq.(\ref{phivarphi}) where
due to a well known relation
$$
\wp'^2 = 4 \wp^3 - g_2 \wp-g_3 \equiv 4(u-e_1)(u-e_2)(u-e_3).
$$
We choose for $dz/du$ the determination
$$
\frac{dz}{du} = -\frac{1}{2\sqrt{(u-e_1)(u-e_2)(u-e_3)}}
$$
where the square root appearing in the denominator is positive for large 
positive $u$.

It is useful to choose the cuts in the
$u$-plane as to have all the source singularities which lay in the fundamental 
region of the $z$-plane mapped onto first sheet of the $u$-plane. 
The shape of such cuts is very
simple when one has one, two or three sources. It becomes in
general more complicated for a higher number of sources but it always exists.

In the present paper we shall restrict to the case in which the sources are
symmetrical w.r.t. the reflection $z\rightarrow -z$. In this case it is not
necessary to distinguish among the two sheets in $u$. We shall devote an other
paper to non symmetric situation. For symmetric sources the problem is
invariant under the transformation $z\rightarrow -z$ because the boundary of
the fundamental region is also invariant under such a transformation. This
implies that also the solution is parity invariant i.e. $\phi(-x,-y) =
\phi(x,y)$. In fact otherwise we shall have at least two solutions of the
problem, while a result by Picard \cite{picard} (see also
\cite{lichtenstein,troyanov})
says that the solution exists and is unique.
Thus for a symmetric distribution of sources the solution will be a single
valued function of $u$.

The standard procedure for solving eq.(\ref{varphiliouville}) is to notice
that as a consequence of 
the relation 
$$
\partial_u\partial_{\bar u} \varphi = c e^{\varphi}
$$
away from the sources we have
$$
\partial_{\bar u}(e^{\frac{\varphi}{2}}\partial^2_u e^{-\frac{\varphi}{2}})=0
$$
i.e. 
$$
e^{\frac{\varphi}{2}}\partial^2_u e^{-\frac{\varphi}{2}}
$$
is an analytic function of $u$, call it $-Q(u)$. Then the general form of
$e^{-\frac{\varphi}{2}}$ is
\begin{equation}\label{emvarphi2}
e^{-\frac{\varphi}{2}} = \bar \chi_1(\bar u) \psi_1(u)+
\bar \chi_2(\bar u) \psi_2(u)
\end{equation}
where $\psi_1$ and $\psi_2$ are two linearly independent solutions of
\begin{equation}\label{fuchsianequation}
y''(u)+Q(u) y(u) =0.\end{equation}
Noticing that we have also
$$
\partial_{u}(e^{\frac{\varphi}{2}}\partial^2_{\bar u} 
e^{-\frac{\varphi}{2}})=0
$$
and due to the reality of $e^{-\frac{\varphi}{2}}$
$$
e^{\frac{\varphi}{2}}\partial^2_{\bar u} e^{-\frac{\varphi}{2}}=
-\bar Q(\bar u)
$$
we find that also $\chi_1$ and $\chi_2$ are solutions of 
eq.(\ref{fuchsianequation}).
Being the $\chi_j$ combinations of the $\psi_j$ again due to the reality of 
$e^{-\frac{\varphi}{2}}$ we can rewrite eq.(\ref{emvarphi2}) in the form
\begin{equation}\label{Hexpression}
e^{-\frac{\varphi}{2}} = \bar \psi_j(\bar u) H_{jk}\psi_k(u)
\end{equation}
with $H_{jk}$ hermitean matrix.
$H$ can always be diagonalized by means of
a unitary transformation to the form $\mu_1 \bar y_1y_1+\mu_2 \bar y_2y_2$,
with $\mu_1,\mu_2$ real and by a further rescaling of the $y_k$ to one of the
forms $\bar y_1 y_1\pm \bar y_2y_2$. The case of both negative eigenvalues
is excluded by the positivity of $e^{-\frac{\varphi}{2}}$.

Denoting with $w_{12}$ the constant Wronskian
$$
w_{12}= y_1 y'_2-y'_1 y_2
$$
it is easily seen by explicit computation that for the upper sign
\cite{redi} away from the sources we have
$$
\partial_u\partial_{\bar u} \varphi= -2|w_{12}|^2 e^{\varphi}
$$ 
while for the lower sign 
$$
\partial_u\partial_{\bar u} \varphi= 2 |w_{12}|^2 e^{\varphi}
$$ 
which is the case we are interested in 
and we need to choose $|w_{12}|^2=1/2$.  Thus we have
reached for the solution of eq.(\ref{varphiliouville}) the form
\begin{equation}\label{RHvarphi}
e^{\varphi}=\frac{2 |w_{12}|^2}{[\bar y_1(\bar u)y_1(u)
-\bar y_2(\bar u) y_2(u)]^2}
\end{equation}
where introducing explicitely the Wronskian we relaxed the condition
$|w_{12}|^2=1/2$. 
The sources appearing in eq.(\ref{varphiliouville}) require the behavior 
of $y_k(u)$ to be
power-like at the singularities. It means that $Q(u)$ has to be a meromorphic 
function
with only first and second order poles i.e. of the form
\begin{eqnarray}
Q(u) &=& \frac{3}{16}\left(\frac{1}{(u-e_1)^2}+\frac{1}{(u-e_2)^2}
       +\frac{1}{(u-e_3)^2}\right)+
\frac{b_1}{2(u-e_1)}+\frac{b_2}{2(u-e_2)}+\frac{b_3}{2(u-e_3)}\nonumber\\
&+& \sum_{n>0}\left(\frac{1-\lambda_n^2}{4(u-u_n)^2}+
\frac{\beta_n}{2(u-u_n)}\right)
\end{eqnarray}
where the $b_k$ and $\beta_n$  are the accessory parameters and they are 
subject to the Fuchs conditions
\begin{equation}\label{bequation}
b_1+b_2+b_3+\sum_{n>0}\beta_n=0
\end{equation}
\begin{equation}\label{beequation}
\frac{9}{16}+\frac{b_1}{2} e_1+\frac{b_2}{2} e_2+\frac{b_3}{2} e_3 
+\sum_{n>0}\frac{\beta_n}{2} u_n+\sum_{n>0}\frac{1-\lambda^2_n}{4}
= \frac{4-\lambda^2}{16}
\end{equation}
where $\lambda = 1-2 \eta_0$  is given by the source at the origin in
eq.(\ref{philiouville}). 
At infinity $Q(u)$ behaves like
$$
\frac{4-\lambda^2}{16 u^2}
$$
and it assures that the $\phi$ field of eq.(\ref{philiouville}) 
obtained from $\varphi$
has the source at the origin $2\pi\delta^2(z)\eta_0$ with $\eta_0= 
(1-\lambda)/2$.

In the case of a single source placed at the origin the general solution of 
eqs.(\ref{bequation},\ref{beequation}) gives for 
$Q(u)$
\begin{eqnarray}
Q(u)&=&
\frac{1-\lambda^2}{16}\frac{u+\beta}{(u-e_1)(u-e_2)(u-e_3)}\\
&+&\frac{3}{16}\left(\frac{1}{(u-e_1)^2}+ \frac{1}{(u-e_2)^2}
+\frac{1}{(u-e_3)^2}+\frac{2 e_1}{(e_1-e_2)(e_3-e_1)(u-e_1)}\right.\nonumber\\
&+& \left.\frac{2 e_2}{(e_2-e_3)(e_1-e_2)(u-e_2)} 
+\frac{2 e_3}{(e_3-e_1)(e_2-e_3)(u-e_3)}\right).\nonumber
\end{eqnarray}
Thus we are left with a single accessory parameter 
$\beta$ which has to be
determined by 
imposing the monodromicity of the solution. We notice that
contrary to the sphere case with three singularities the simple Fuchs 
conditions leave one accessory parameter undetermined. 

For the $\varphi$ appearing in eq.(\ref{RHvarphi}) to be solution of
eq.(\ref{varphiliouville}) it must be monodromic around the singularities for
which it is necessary and sufficient that under a tour around each singularity
the pair of function $y_1,y_2$ undergoes an $SU(1,1)$ transformation.
  
\section{The periods of the torus}\label{periods}

In this section we summarize some facts about the periods of the torus
\cite{batemanII} in relation to the roots of the equation  
\begin{equation}\label{gequation}
4u^3 - g_2 u -g_3=4(u-e_1)(u-e_2)(u-e_3)
\end{equation}
and what happens in the particular cases in which the fundamental region 
becomes a rectangle or a square. Without loss of generality we can set
$\omega_1 ={\rm real}$, $\omega_1>0$  and $\omega_2 = \omega_2^R + 
i \omega_2^I$ with
$\omega_2^I >0$.
From eq.(\ref{gequation}) we have
\begin{equation}\label{esumrule}
e_1+e_2+e_3=0.
\end{equation}
In the case of the rectangle we have $\omega_1 = \omega_1^R$ and $\omega_2=
i\omega_2^I$  and as a consequence
\begin{equation}\label{g2}
g_2 \equiv 60{\sum_{mn}}'\frac{1}{(2m\omega_1+2n\omega_2)^4}={\rm real}
\end{equation}
\begin{equation}\label{g3}
g_3 \equiv 140{\sum_{mn}}'\frac{1}{(2m\omega_1+2n\omega_2)^6}={\rm real}
\end{equation}
where in the sums the term $m=n=0$ is excluded.
The discriminant of eq.(\ref{gequation}) is
$$
\Delta = g_2^3 -27 g_3^2 = 16(e_1-e_2)^2(e_2-e_3)^2(e_3-e_1)^2
$$
and we have \cite{WW}
\begin{equation}\label{DeltaWW}
\Delta =\wp'(z)\wp'(z+\omega_1)\wp'(z+\omega_2)
\wp'(z+\omega_1+\omega_2).
\end{equation}
As for $\omega_1$ real and $\omega_2$ pure imaginary the $\wp$ function
is a real analytic function i.e. $\wp(\bar z)= \overline{\wp(z)}$
we have also $\wp'(\bar z)= \overline{\wp'(z)}$. Choosing in
eq.(\ref{DeltaWW}) $z= -\frac{\omega_1}{2}-\frac{\omega_2}{2}$ we reach the
conclusion $\Delta \geq 0$. As $g_2$ and $g_3$ are real we have that one of
the solution, say $e_2$ is real. Then if $e_1$ has a non zero imaginary part
we have $e_3= \bar e_1$ and
\begin{equation}
\Delta = 16(e_2-e_1)^2 (e_2-\bar e_1)^2 (2i {\rm Im}(e_1))^2 <0
\end{equation}
which violates $\Delta \geq 0$. Thus we have $e_1$ real and as a consequence
$e_3$ real. The conclusion is that for the rectangle we have all $e_j$ real.
In the case of the square i.e. $\omega_2 = i \omega_1$ with $\omega_1$ real,
it is easily shown from the expression (\ref{g3}) that $g_3=0$. Then
say $e_2=0$ and as $e_1+e_2+e_3=0$ we have $e_3 = -e_1$ and as we are in a
special case of rectangle we also have $e_1$ real and $e_3=-e_1$.

The inverse function of $u=\wp(z)$ can be written as
$$
z(u) = \int^\infty_u\frac{dx}{2\sqrt{(x-e_1)(x-e_2)(x-e_3)}}~.
$$
For large positive $x$ the square root is defined as $+(x)^{3/2}$. 
As $z$ goes along
the path $0,\omega_1$, $\omega_1+\omega_2\equiv \omega_3$, $\omega_2$, 
$u$ follows a path from $+\infty$, $e_1$, $e_3$ and $e_2$. 
When $u$  reaches $e_1$, $u$ must continue below the
branch point so that $z$ moves upward in the complex plane and again when $u$
reaches $e_3$ it must continue below the branch point as to have $z$ moving
to the left in the complex plane.

\section{Perturbative solution for the one point function}
\label{pertonepoint}

In this section we shall consider the case of a single weak source at the
origin, i.e. $\eta_0 = (1-\lambda)/2$ small. One can similarly apply the
technique to several weak sources.

For $\lambda =1$ it is possible to
solve eq.(\ref{fuchsianequation}) exactly. Two independent solutions are 
given by
\begin{equation} 
y_1(u)  = [(u-e_1)(u-e_2)(u-e_3)]^{\frac{1}{4}}\equiv \Pi(u) 
\end{equation}
\begin{equation}
y_2(u) = \Pi(u) Z(u)
\end{equation}
where
\begin{equation}
Z(u) \equiv z(u) - \omega_1-\omega_2=z(u) - \omega_3
\end{equation}
so that $Z(e_1) = - \omega_2$, $Z(e_3) =0$, $Z(e_2) = - \omega_1$.

The Wronskian is given by
\begin{equation}
w_{12}=y_1y'_2-y'_1y_2= \Pi^2(u) Z'(u)=
-\frac{1}{2} 
\end{equation}
where the determination of $\Pi(u)$ has been chosen so that
$$
\Pi^2(u) = \sqrt{(u-e_1)(u-e_2)(u-e_3)}~.
$$
We shall solve eq.(\ref{fuchsianequation}) perturbatively by writing the 
solution as
$y_j+\delta y_j$ 
$$
\delta y_j'' + Q_0 \delta y_j = - q y_j
$$
where
\begin{eqnarray}
Q_0(u)&=&\frac{3}{16}\left(\frac{1}{(u-e_1)^2}+ \frac{1}{(u-e_2)^2}+
\frac{1}{(u-e_3)^2}+\frac{2 e_1}{(e_1-e_2)(e_3-e_1)(u-e_1)}\right.\nonumber\\
&+& \left.\frac{2 e_2}{(e_2-e_3)(e_1-e_2)(u-e_2)} 
+\frac{2 e_3}{(e_3-e_1)(e_2-e_3)(u-e_3)}\right)\nonumber\\
\end{eqnarray}
and $q(u)$
$$
q(u) = \frac{\varepsilon}{16}\frac{u+\beta}{(u-e_1)(u-e_2)(u-e_3)}
$$
with $\varepsilon = 1-\lambda^2$.
The Green function of the unperturbed equation is
\begin{equation}
G(u,x) = \frac{1}{w_{12}}\Big(\Theta(u,x) y_1(u) y_2(x)-
\Theta(u,x) y_2(u) y_1(x) \Big).
\end{equation}
The perturbed solutions are
\begin{equation}
y_j +\delta y_j 
\end{equation}
with
\begin{eqnarray}\label{deltaygeneral}
& &\delta y_j(u) = \int_{e_3}^{\infty} G(u,x) q(x) y_j(x) dx=\\
& &=\frac{1}{w_{12}}\left(y_1(u)\int^u_{e_3} y_2(x) q(x) y_j(x) dx -
y_2(u)\int^u_{e_3} y_1(x) q(x) y_j(x) dx \right)\\
& &=\frac{1}{w_{12}}\left(y_1(u) I_{j2}(u) -
y_2(u) I_{j1}(u)\right)
\end{eqnarray}
with
\begin{equation}
I_{jk}(u) = \int^u_{e_3} q(x) y_j(x)y_k(x) dx.
\end{equation}
Thus
$$
I_{11}(u) = \frac{\varepsilon}{16}\int_{e_3}^u\frac{(x+\beta)dx}
{\sqrt{(x-e_1)(x-e_2)(x-e_3)}}
$$
$$
I_{12}(u) = \frac{\varepsilon}{16}\int_{e_3}^u\frac{(x+\beta)Z(x)dx}
{\sqrt{(x-e_1)(x-e_2)(x-e_3)}}~,
$$
$$
I_{22}(u) = \frac{\varepsilon}{16}\int_{e_3}^u\frac{(x+\beta)Z^2(x)dx}
{\sqrt{(x-e_1)(x-e_2)(x-e_3)}}.
$$
The accessory parameter $\beta$ will have to be fixed as to have that for a
tour around each singularity $e_k$ and infinity the $\varphi$ appearing in
eq.(\ref{RHvarphi}) be monodromic.

At $e_3$ the behavior of the $I_{jk}$ is the following
$$
I_{11}(u) = (u-e_3)^{\frac{1}{2}} f_{11}(u-e_3),~~~~
I_{12}(u) = (u-e_3) f_{12}(u-e_3),~~~~
I_{22}(u) = (u-e_3)^{\frac{3}{2}} f_{22}(u-e_3)
$$
where $f_{jk}$ and the $f_j$ immediately below denote power series in 
$u-e_3$. Combining with the behavior of $y_j$ at $e_3$ we have that
$$
\delta y_1 = (u-e_3)^{\frac{5}{4}} f_1(u-e_3),~~~~ \delta y_2 =
(u-e_3)^{\frac{7}{4}} f_2(u-e_3)
$$
so that the monodromic behavior of the perturbed functions at $e_3$ is
unchanged independently of the value of $\beta$.
Notice that the Wronskian $w_{12}$ is left unchanged
by the perturbation. 

We examine now the monodromic properties at $e_1$. With regard to $y_1
=\Pi(u)$ the expression $\bar y_1 y_1$ is monodromic around $e_1$.  Instead
for $y_2=\Pi(u)Z(u)$ we have for $Z(u)$
around $e_1$
\begin{eqnarray}
Z(u) &\approx& \frac{1}{2}\int_{e_1}^{e_3} 
\frac{dx}{\sqrt{(x-e_1)(x-e_2)(x-e_3)}} -
\frac{(u-e_1)^{\frac{1}{2}}}{\sqrt{(e_1-e_2)(e_1-e_3)}}=\\
&=& -\omega_2-\frac{(u-e_1)^{\frac{1}{2}}}{\sqrt{(e_1-e_2)(e_1-e_3)}}.
\end{eqnarray}
Then 
\begin{equation}
y_2 \approx \Pi(u)[-\omega_2 -s_1 (u-e_1)^{\frac{1}{2}}]
\end{equation}
where
\begin{equation}
s_1= \frac{1}{\sqrt{(e_1-e_2)(e_1-e_3)}}.
\end{equation}
Thus $\bar y_2 y_2$ is not monodromic at $e_1$.

We examine now $\delta y_1$ around $e_1$.
\begin{eqnarray}
\delta y_1 &=& \frac{\Pi(u)}{w_{12}}
\left[I_{12}(e_1)+\frac{\varepsilon}{16}\int_{e_1}^u\frac{(x+\beta)Z(x)dx}
{\sqrt{(x-e_1)(x-e_2)(x-e_3)}}\right.\nonumber\\
&-&
\left.Z(u)\frac{\varepsilon}{16}\int_{e_3}^u\frac{(x+\beta)dx}
{\sqrt{(x-e_1)(x-e_2)(x-e_3)}}\right]\nonumber\\
&\approx& \frac{\Pi(u)}
{w_{12}}
\left[I_{12}(e_1)+\frac{\varepsilon}{16}~2s_1(e_1+\beta)Z(e_1)
(u-e_1)^{\frac{1}{2}}\right.
\nonumber\\
&-&\left.(Z(e_1)-s_1(u-e_1)^{\frac{1}{2}})
(I_{11}(e_1)+\frac{\varepsilon}{16}~
2s_1(e_1+\beta)(u-e_1)^{\frac{1}{2}})\right]\nonumber\\
\end{eqnarray}
i.e.
$$
y_1+\delta y_1 =\Pi(u)\left[1+ \varepsilon~{\rm const}~+\frac{s_1I_{11}(e_1)
(u-e_1)^{\frac{1}{2}}}{w_{12}}\right].
$$
Then we must impose the monodromicity of
\begin{equation}
\left|\Pi(u)\left(1+\varepsilon~{\rm const}~+
\frac{s_1 I_{11}(e_1) (u-e_1)^{\frac{1}{2}}}{w_{12}}\right) \right|^2-
|k|^2 \left|\Pi(u)\left(-\omega_2-s_1 (u-e_1)^{\frac{1}{2}}\right)\right|^2 
\end{equation}            
where being already $|k|^2$ of first order in $\varepsilon$ we have neglected
the correction $\delta y_2$ to $y_2$.

To order $\varepsilon$, taking into account that the $I_{jk}$ are themselves of
order $\varepsilon$, the monodromicity of the previous expression becomes
\begin{equation}
0=\frac{I_{11}(e_1)s_1(u-e_1)^{\frac{1}{2}}}{w_{12}}
-|k|^2 \bar \omega_2 s_1(u-e_1)^{\frac{1}{2}}
\end{equation}
or
\begin{equation}\label{k2e1}
0<|k|^2 =\frac{I_{11}(e_1)}{w_{12} \bar \omega_2}.
\end{equation}
$I_{11}(e_1)$ can be explicitely computed because we have
\begin{eqnarray}
I_{11}(e_1)&=&\frac{\varepsilon}{16}
\int_{e_3}^{e_1}\frac{(\beta+x)dx}{\sqrt{(x-e_1)(x-e_2)(x-e_3)}}=
\frac{\varepsilon}{8}
\left[\beta (z(e_3)-z(e_1)) -\int_{z(e_3)}^{z(e_1)}\wp(z)dz\right]
\nonumber \\
&=& \frac{\varepsilon}{8}\left[\beta \omega_2
+\zeta(\omega_1)-\zeta(\omega_1+\omega_2)\right]
=\frac{\varepsilon}{8}\left[\beta \omega_2 -\zeta(\omega_2)\right]
\end{eqnarray}
where we used the property of the $\zeta$ function \cite{batemanII}
\begin{equation}\label{zetaprime}
\zeta'(z) = -\wp(z) 
\end{equation}
and we know that \cite{batemanII}
$$
\zeta(z+ 2\omega_j) = \zeta(z)+2\zeta(\omega_j),~~~~
\zeta(\omega_1+\omega_2) = \zeta(\omega_1) +\zeta(\omega_2). 
$$
Similarly we have, working around the point $e_2$
\begin{equation}\label{k2e3}
0<|k|^2 =\frac{I_{11}(e_2)}{w_{12} \bar \omega_1}
\end{equation}
with
$$
I_{11}(e_2) = \frac{\varepsilon}{8}[\beta \omega_1 -\zeta(\omega_1)]. 
$$
Solving the system of eqs.(\ref{k2e1},\ref{k2e3})
we have
$$
\beta = \frac{\zeta(\omega_2)\bar \omega_1 - \zeta(\omega_1)\bar
\omega_2}{\omega_2 \bar \omega_1 - \omega_1\bar\omega_2} 
$$
and
$$
|k|^2 
= \frac{\varepsilon(\omega_2\zeta(\omega_1)-\omega_1\zeta(\omega_2))}
{8w_{12}(\omega_1\bar \omega_2-\omega_2\bar \omega_1)}~.
$$
Using now Legendre relation \cite{batemanII}
$$
\omega_2\zeta(\omega_1) -\omega_1\zeta(\omega_2) =\frac{\pi i}{2} 
$$
which holds for ${\rm Im}(\frac{\omega_2}{\omega_1})>0$ which is our case,
we have
$$
|k|^2 = \frac{\varepsilon\pi}{4 A}
$$
being $A$ the area of the fundamental region. We see that such a relation can
be satisfied with $|k|^2 >0$ only for $\varepsilon>0$ i.e. $\eta>0$ in
accordance with the Gauss-Bonnet theorem.
The quadratures appearing in $\delta y_1(u)$ i.e. the $I_{12}(u), 
~I_{11}(u)$ 
can all 
be done in terms of elliptic functions. Using eq.(\ref{zetaprime}) and
\cite{batemanII} 
$$
\zeta(z)
=\frac{\zeta(\omega_1)}{\omega_1}z+
\frac{\theta'_1(v|\tau)}
{2\omega_1\theta_1(v|\tau)},~~~~{\rm with}~~v= \frac{z}{2\omega_1}~~{\rm and}~~
\tau =\frac{\omega_2}{\omega_1} 
$$
we obtain, making use again of the Legendre relation,
\begin{eqnarray}
&&e^{-\frac{\phi}{2}}=
\frac{1}{\sqrt{2}~|k|}\left(1+\frac{\varepsilon}{4}\left[
\left(
\log\frac{\theta_1(\frac{z}{2\omega_1}|\tau)}{\theta_1(\frac{\omega_3}
{2\omega_1}|\tau)}+
c.c.\right)- \right.\right.
\\
& &\left.\left.\frac{i\pi}{4(\omega_1\bar\omega_2-\bar\omega_1\omega_2)}\left(
\frac{\bar\omega_1}{\omega_1} z^2+\frac{\omega_1}{\bar\omega_1} \bar
z^2-2z\bar z\right)+\frac{i\pi}{\omega_1\bar\omega_1}
(\omega_1\bar\omega_2-\bar\omega_1\omega_2)\right]\right).
\end{eqnarray}
The derivative of $\phi$ with respect to $\varepsilon$ obviously provides the
Green function on the flat torus, which is defined up to an additive
constant. 
One obtains with the standard values 
$2\omega_1=1$ and $2\omega_2 = \tau^R+i\tau^I$
\begin{equation}\label{Greenfunction}
G(z) = \frac{1}{4\pi}\log|\theta_1(z|\tau)|^2 +\frac{1}{8\tau^I}(z-\bar z)^2
\end{equation}
satisfying
$$
\Delta G(z) = \delta(z) -\frac{1}{\tau^I}.
$$
Equation (\ref{Greenfunction}) was derived in 
\cite{itzykson,difrancesco,weil} by summing the
Fourier representation  of the Green function in momentum space by using
Kronecker resummation formula \cite{weil}. Here it has been derived by a 
quadrature.

\section{Counting the degrees of freedom}\label{degreesoffreedom}

The above perturbative calculation raises the following general question which
we want to investigate at the non perturbative level. We have to satisfy the
monodromy of $\varphi$ at $e_1$, $e_2$ having at our disposal only the three 
real parameters $\beta^R, \beta^I$ and $|k|^2$ the last of which 
has to result positive. Monodromy of $\varphi$ means that the monodromy
matrices $M(e_k)$ at 
$e_1$ and $e_2$ belong to $SU(1,1)$, being this a necessary and
sufficient condition for monodromy and three real parameters appear too few to 
satisfy the four 
complex equations for the matrix elements 
$M_{22}(e_k) = \overline {M}_{11}(e_k)$,  
$M_{21}(e_k) = \overline{M}_{12}(e_k)$  with $k=1,2$.

We choose as before $y_k$ to be canonical at $e_3$
i.e. $y_1 = (u-e_3)^{\frac{1}{4}}f_1(u-e_3),~~y_2 =
(u-e_3)^{\frac{3}{4}}f_2(u-e_3)$. 
If we denote by $S$ the $SL(2,C)$ transformation
which expresses the solution canonical at $e_3$
$$
Y=
\begin{pmatrix}
y_1\\
y_2
\end{pmatrix}
$$
around $e_1$ as
$$
Y= S
\begin{pmatrix}
\zeta^{\frac{1}{4}}g_1(\zeta)\\
\zeta^{\frac{3}{4}}g_2(\zeta)
\end{pmatrix}
$$
with $\zeta = u-e_1$
and
$$
S=
\begin{pmatrix}
a&b\\
c&d
\end{pmatrix}
$$
we have for the monodromy matrix at $e_1$
$$
M=i
\begin{pmatrix}
ad+bc&-2ab\\
2cd&-ad-bc
\end{pmatrix}.
$$
We can also multiply $Y$ by the matrix
$$
K=
\begin{pmatrix}
\kappa^{-1}&0\\
0&\kappa
\end{pmatrix}
$$
without altering its monodromy property at $e_3$. Then we find for the
monodromy matrix at $e_1$
$$
M=
\begin{pmatrix}
i(ad+bc)&-2iab\kappa^{-2}\\
2icd\kappa^2&- i(ad+bc)
\end{pmatrix}
$$
i.e. we have at all the points $e_k$
\begin{equation}\label{Mproperties}
1 = M_{11} M_{22} -M_{12} M_{21}~~~~{\rm and}~~~~M_{11}+M_{22}=0
\end{equation}
independently of the imposition of the monodromy. 

Imposition of the $SU(1,1)$ nature of $M$ gives
$$
ab \kappa^{-2}=\bar c\bar d  \bar\kappa^{2}~~~~{\rm i.e.}~~~~
\kappa^2\bar \kappa^2 = \frac{ab}{\bar c\bar d}~.
$$
This is sufficient also to give $M_{22}= \overline {M_{11}}$ because we have
$$
abcd={\rm real}>0,~~~~{\rm i.e.}~~~ad =
r_1\theta,~~~~bc=r_2\bar\theta,~~~~r_1\geq 0,~~r_2 \geq 0 
$$
with $\theta$ a unitary number, which combined with
$$
ad-bc=1 =r_1\theta-r_2\bar\theta
$$
gives
$$
(r_1+r_2)(\theta-\bar\theta)=0,~~~~{\rm i.e.}~~~~ad={\rm real},
~~~~bc={\rm real}.
$$
We can spend now the complex
parameter $\beta=\beta^R+i\beta^I$ for solving the complex equation 
\begin{equation}\label{abcdratios}
\frac{ab}{\bar c\bar d}(e_1)= \frac{ab}{\bar c\bar d}(e_2)
\end{equation}
and we are left with proving the reality and positivity of such a ratio.

At the perturbative level we saw that Legendre relation solves this problem by
giving to $\kappa^2\bar\kappa^2$ a real positive value as it must be. 
We want to understand here this problem at the nonperturbative level.

The monodromy matrix at $e_3$ where the solutions are canonical with indices
$1/4,~3/4$ is given by 
\begin{equation}
M(e_3)=D(0)=
\begin{pmatrix}
i& 0\\
0&-i
\end{pmatrix},
\end{equation}
and we have the group relation
\begin{equation}
D(0) M(e_2) M(e_1) = M^{-1}(\infty)
\end{equation}
and by construction all matrices are of $SL(2,C)$ type. Using 
eq.(\ref{Mproperties}) we can write
\begin{equation}
M(e_2)=
\begin{pmatrix}
m_{11}& r_1e^{i\phi_1}\\
r_2e^{i\phi_2}&-m_{11}
\end{pmatrix},~~~~
M(e_1)=
\begin{pmatrix}
n_{11}&s_1e^{i\psi_1}\\
s_2e^{i\psi_2}&-n_{11}
\end{pmatrix}.
\end{equation}
We have
\begin{equation}\label{trace}
{\rm Tr}\left(D(0) M(e_2) M(e_1)\right) 
= i [\left(M(e_2)M(e_1)\right)_{11}-\left(M(e_2)M(e_1)\right)_{22}]=
- 2 \cos\frac{\pi\lambda}{2} = {\rm real}
\end{equation}
which becomes
\begin{equation}
i(r_1s_2 e^{i(\phi_1+\psi_2)}-r_2s_1
e^{i(\phi_2+\psi_1)}) =\rm{real}
\end{equation}
or
\begin{equation}\label{cosineequation}
r_1s_2 \cos(\phi_1+\psi_2)-r_2s_1 \cos(\phi_2+\psi_1)=0. 
\end{equation}
As already mentioned we can spend $\beta^R+i\beta^I$ 
to satisfy eq.(\ref{abcdratios}) i.e. $r_1/r_2=s_1/s_2$ and 
$e^{i(\phi_1+\phi_2)}=
e^{i(\psi_1+\psi_2)}$. 
Thus
\begin{equation}\label{phaseequality}
\phi_1+\phi_2 = \psi_1+\psi_2~({\rm mod}~2\pi).
\end{equation}
We cannot have for the solution of (\ref{cosineequation})
\begin{equation}
\phi_1+\psi_2=\psi_1+\phi_2 ~({\rm mod}~2\pi)
\end{equation}
otherwise $\rm{Tr} M(\infty)=0$ for the $\beta$ which realizes
eq.(\ref{abcdratios}) against eq.(\ref{trace}).
Then
\begin{equation}
\phi_1+\psi_2 = -\psi_1-\phi_2~~({\rm mod}~2\pi), ~~~~\rm{or}~~~~
\phi_1+\phi_2 = -\psi_1-\psi_2 ~~({\rm mod}~2\pi)
\end{equation}
which combined with (\ref{phaseequality}) gives
\begin{equation}
\psi_2 = -\psi_1~({\rm mod}~2\pi),~~~~~~~~\phi_2 = -\phi_1~({\rm mod}~2\pi)
\end{equation}
and we have a $\kappa$ 
which renders $M(e_3)$, $M(e_2)$, $M(e_1)$, and thus also $M(\infty)$
of $SU(1,1)$ type.
The results of \cite{picard,lichtenstein,troyanov} 
tell us that when Picard's inequalities 
are satisfied, 
such $\beta$ exists in agreement with the perturbative result.

\section{The square}\label{square}

We shall now solve at the non perturbative level a special case, i.e. the one
in which the fundamental region of the torus is a square. In such a situation
the function $Q(u)$ becomes with $e_3=0, e_1=-e_2=1$ (see Sec.\ref{periods}) 
\begin{equation}
Q(u) = \frac{(1-\lambda^2)(u+\beta)}{16u(u^2-1)}+
\frac{3}{16}\frac{(1+u^2)^2}{u^2(1-u^2)^2}~.
\end{equation}
The following formal
argument supports the result $\beta=0$ for the square. Suppose the monodromy is
achieved for a certain value of $\beta$. We can now perform the transformation
$u\rightarrow -u$ which is equivalent to $\beta\rightarrow -\beta$. Thus if
$\beta$ is a solution  also $-\beta$ is, and if the solution is unique we have
$\beta=0$. We shall prove explicitely in the following that the differential
equation with $\beta=0$ solves all the monodromy conditions.

Setting $\beta=0$ we have
\begin{equation}
Q(u) = \frac{1-\lambda^2}{16(u^2-1)}+\frac{3}{16}\frac{(1+u^2)^2}{u^2(1-u^2)^2}
\end{equation}

Using $x=u^2$ we have
\begin{equation}
4x\frac{d^2y}{dx^2}+2\frac{dy}{dx}+\left[\frac{1-\lambda^2}{16(x-1)}+
\frac{3}{16}\frac{(1+x)^2}{x(1-x)^2}\right]y=0  
\end{equation} 
giving rise to the following $P$-symbol
\begin{equation}
P
\begin{pmatrix}
0 &\infty&1\\
3/8&-1/4+\lambda/8&3/4 &x\\
1/8&-1/4-\lambda/8 &1/4
\end{pmatrix}.
\end{equation}
Two independent solutions canonical at $e_3=0$ are
\begin{eqnarray}\label{y1y2hyper}
y_1&
=&u^{\frac{1}{4}}(1-u^2)^{\frac{1}{4}}
F(\frac{1+\lambda}{8},\frac{1-\lambda}{8};\frac{3}{4};u^2)\nonumber\\
y_2&=&u^{\frac{3}{4}}(1-u^2)^{\frac{1}{4}}
F(\frac{3+\lambda}{8},\frac{3-\lambda}{8};\frac{5}{4};u^2).
\end{eqnarray}
The analytic continuation at $u^2=1$ is given by \cite{batemanII}
\begin{eqnarray}
y_1 = u^{\frac{1}{4}}(1-u^2)^{\frac{1}{4}}[A_1 F(a,b;a+b-c+1;1-u^2)+
\nonumber\\
A_2 (1-u^2)^{c-a-b}F(c-a,c-b;c-a-b+1;1-u^2)]
\end{eqnarray}
\begin{eqnarray}
y_2 = u^{\frac{3}{4}}(1-u^2)^{\frac{1}{4}}[A'_1 F(a',b';a'+b'-c'+1;1-u^2)+
\nonumber\\
A'_2 (1-u^2)^{c'-a'-b'}F(c'-a',c'-b';c'-a'-b'+1;1-u^2)]
\end{eqnarray}
with
\begin{equation}
A_1 =\frac{\Gamma(c)\Gamma(c-a-b)}{\Gamma(c-a)\Gamma(c-b)},~~~~
A_2 =\frac{\Gamma(c)\Gamma(a+b-c)}{\Gamma(a)\Gamma(b)}
\end{equation}
giving
\begin{equation}\label{kappa4square}
|\kappa|^4 =\frac{ab}{\bar c \bar d}=
\left(\frac{\Gamma(3/4)}{\Gamma(5/4)}\right)^2 
~\frac{\Gamma(\frac{3+\lambda}{8})
\Gamma(\frac{3-\lambda}{8}) \Gamma(\frac{7+\lambda}{8})
\Gamma(\frac{7-\lambda}{8})}
{\Gamma(\frac{1-\lambda}{8})
\Gamma(\frac{1+\lambda}{8}) \Gamma(\frac{5+\lambda}{8})
\Gamma(\frac{5-\lambda}{8})} ~.
\end{equation}

This is the non perturbative value of $|\kappa|^4$; for small $\varepsilon 
\equiv 1-\lambda^2 \approx 2(1-\lambda)$ we have
\begin{equation}\label{kappa5square}
|\kappa|^4 = \frac{ab}{\bar c \bar d}\approx 
\left(\frac{\Gamma(3/4)}{\Gamma(5/4)}\right)^2 
\frac{1}{\Gamma(\frac{\varepsilon}{16})}
\approx\left(\frac{\Gamma(3/4)}{\Gamma(5/4)}\right)^2 \frac{\varepsilon}{16}~.
\end{equation}
This is in agreement with the perturbative calculation given previously 
$$
|\kappa|^4 = \frac{\varepsilon\pi}{4 A}
$$
as in our case we have $g_2=4, g_3=0$ and
$$
\omega_1 = -i \omega_2
=\frac{1}{2}\int_1^\infty\frac{du}{\sqrt{u^3-u}}=\frac{\sqrt{\pi}\Gamma(5/4)}
{\Gamma(3/4)}=1.31103... 
$$
Taking into account eqs.(\ref{phivarphi},\ref{RHvarphi}) we have for the 
square torus
\begin{equation}\label{squarephi}
e^{-\frac{\phi(z)}{2}}=\frac{1}{\sqrt{2}|\kappa|^2}
\left[\left|F(\frac{1+\lambda}{8},\frac{1-\lambda}{8};\frac{3}{4};u^2(z))
\right|^2- |\kappa|^4 |u(z)|
\left|F(\frac{3+\lambda}{8},\frac{3-\lambda}{8};\frac{5}{4};u^2(z))\right|^{2}
\right]
\end{equation}
where $|\kappa|^4$ is given by eq.(\ref{kappa4square}).

From  the derivative of $\phi$, given by eq.(\ref{squarephi}),
 with respect to $\varepsilon$ for $\varepsilon=0$ we obtain the Green 
function for the torus given by the square with half periods
$\omega_1 = - i \omega_2$ related to the invariants we are working with i.e. 
$g_2=4,~g_3=0$  ($\omega_1 =-i \omega_2 = 1.31103..$.)
$$
G_s(z)= \frac{1}{16\pi}\left(2 ~
{\rm Re}\left(F^{(1)}(0,\frac{1}{4};\frac{3}{4};u^2(z))\right)-
\left(\frac{\Gamma(\frac{3}{4})}
{\Gamma(\frac{5}{4})}\right)^2|u(z)|
\left|F(\frac{1}{4},\frac{1}{2};\frac{5}{4};u^2(z))\right|^2\right)
$$
where $F^{(1)}$ denotes the derivative of the 
hypergeometric function
 with respect to the first argument.
It gives an expression for the Green function on the square alternative to
 eq.(\ref{Greenfunction}). 
The relation of $G_s$ to eq.(\ref{Greenfunction}) is
$$
G_s(2\omega_1 z) = \frac{1}{4\pi}\log|\theta_1(z|i)|^2
 +\frac{1}{8}(z-\bar z)^2 +{\rm const.} 
$$

\section{General deformation}\label{generaldeformation}

We shall develop here the formulas for a deformation starting from a general
configuration; later we shall apply it to the case of the square.

To the half-periods $\omega_1,\omega_2$ there corresponds according to 
the formulas reported in Sec.\ref{periods} the roots 
$e_1,e_2,e_3$ subject to 
the restriction (\ref{esumrule}). It is however simpler in working with 
the differential equation to
work with singularities whose locations are not restricted by 
(\ref{esumrule}). 
The original situation can be recovered by a simple translation.

We shall choose as position of the singularities in the $u$-plane 
$-1,0$ and $u_1$. A general change in $u_1$ has two real degrees of freedom 
which correspond to the changes in $\omega_2/\omega_1$. 
The
remaining transformations in the periods are trivial i.e. the rotations and the
dilatations. Again we shall choose two unperturbed solutions which are
canonical at $0$ and then impose the monodromy requirements in $-1$ and
$u_1$. The first problem is rather simple. The maintenance of the monodromy at
the moving singularity $u_1$ is rather complicated and we shall avoid it by a
proper coordinate transformation.

The $Q$ becomes
\begin{eqnarray}
Q(u) &=& \frac{h}{16}\frac{u+\beta}{u(u+1)(u-u_1)}+\\
& &\frac{3}{16}\left(\frac{1}{(u-u_1)^2}+\frac{1}{u^2}+\frac{1}{(u+1)^2}-
\frac{2}{(u_1+1)(u-u_1)}+\frac{2}{(u_1+1)(u+1)}\right)\nonumber
\end{eqnarray}
and under the perturbation $u_1\rightarrow u_1+\varepsilon$ we have the
variation of $Q(u)$
\begin{eqnarray}
q(u) &=& \varepsilon\left[\frac{h(u^2-\beta u_1)+u(6+\beta h+6u_1-hu_1)}
{16 u(1+u)(u-u_1)^3}\right]+\nonumber \\
&&\beta_1\frac{h}{16u(u+1)(u-u_1)}.
\end{eqnarray}
We now denote with $y_1(u)$, $y_2(u)$ the two solutions, canonical at $u=0$
which behaves respectively as $u^{1/4}$ and $u^{3/4}$ whose Wronskian is
$1/2$. 
The behavior of $y_k(u)$ around $u=-1$ will be
$$
y_1(u) \approx a \zeta^{1/4}+b \zeta^{3/4}
$$
$$
y_2(u) \approx c \zeta^{1/4}+d \zeta^{3/4}
$$
with $\zeta = u+1$ and
$$
\begin{pmatrix}
a &b\\
c&d
\end{pmatrix}
$$
an $SL(2,C)$ matrix.
Using the same notation as in  Sec.\ref{pertonepoint} for the variation 
of the 
solutions we have the behavior around $u=-1$
\begin{eqnarray}\label{gendeltay1}
\delta y_1(u) &\approx& \frac{a\zeta^{1/4}+b\zeta^{3/4}}{w_{12}}\left[
I_{12}(-1)+\int_{-1}^u q(x)y_1(x) y_2(x)dx\right]\nonumber\\
&-&\frac{c\zeta^{1/4}+d\zeta^{3/4}}{w_{12}}\left[
I_{11}(-1)+\int_{-1}^u q(x)y_1(x) y_1(x)dx\right]\approx\\
&&
\frac{a\zeta^{1/4}+b\zeta^{3/4}}{w_{12}} I_{12}(-1)-
\frac{c\zeta^{1/4}+d\zeta^{3/4}}{w_{12}}I_{11}(-1)\nonumber
\end{eqnarray}
the reason being that the terms of order $\zeta^{1/2}$ generated in the
integrals from $-1$ to $u$ cancel, 
leaving only the last line
in eq.(\ref{gendeltay1}) multiplied by terms which are analytic 
in $\zeta$ around $\zeta
=0$. 
Similarly we find
$$
\delta y_2(u) \approx\frac{a\zeta^{1/4}+b\zeta^{3/4}}{w_{12}} I_{22}(-1)-
\frac{c\zeta^{1/4}+d\zeta^{3/4}}{w_{12}}I_{12}(-1).
$$
Thus the perturbed functions near a $u=-1$ are given by 
$(1+F)\begin{pmatrix}
y_1 \\ y_2\end{pmatrix}$ 
where 
$$
F = \frac{1}{w_{12}}
\begin{pmatrix}
I_{12}(-1) & -I_{11}(-1)\\
I_{22}(-1) & -I_{12}(-1)
\end{pmatrix}.
$$
We come now to the imposition of the $SU(1,1)$ nature of the perturbed
monodromy matrix which is the
necessary and sufficient condition for having monodromy around
$u=-1$. Allowing for a multiplication by $(1+\delta K)K$ with
$$
K= 
\begin{pmatrix}
\kappa^{-1}& 0\\
0 & \kappa
\end{pmatrix},~~~~
\delta K =
\begin{pmatrix}
-\delta\kappa& 0\\
0 & \delta\kappa
\end{pmatrix}
$$
being $K$ the unperturbed scale transformation which realized the 
monodromies at all singularities, 
we have around $-1$
$$
Y = (1+\delta K) K(1+F)
\begin{pmatrix}
y_1\\
y_2
\end{pmatrix}
$$
giving for the monodromy matrix
$$
M +[\delta K, M] +[KFK^{-1}, M]
$$
being $M$ the unperturbed monodromy matrix. Thus $\delta M$ is given by
$$
\delta M = [\delta K,M]+[KFK^{-1},M].
$$
Imposition of $\delta M_{21}=\overline{\delta M_{12}}$ gives
$$
w_{12}(\delta\kappa+\delta\bar \kappa) = 
I_{12}(-1)+\overline{ I_{12}(-1)}
+\frac{\overline{M_{11}(-1)}}{\overline{M_{12}(-1)}}
(\overline {\kappa^{-2}} \overline
{I_{11}(-1)} +\kappa^2 I_{22}(-1))
$$
where we took into account the $SU(1,1)$ nature of $M$ and its tracelessness.
Using without loss of generality $\kappa$ and $\delta\kappa$ real we have
\begin{equation}\label{kfromI}
2 w_{12}~\delta\kappa =  I_{12}(-1)+\overline
{I_{12}(-1)}+\frac{\overline{M_{11}(-1)}}{\overline{M_{12}(-1)}}
(\kappa^{-2} 
\overline{I_{11}(-1)} +\kappa^2 I_{22}(-1)).
\end{equation}

We come now to the monodromy at $u_1$. It is cumbersome to impose the
monodromy on a moving singularity even if in principle one can deal with it.
It is much simpler  to perform the following coordinate transformation
$$
u = r\frac{u_1+\varepsilon}{u_1}
$$
and set $t_k(r) = y_k(u)$. The function $t(r)$ obeys the following
differential equation
$$
t''(r) + (Q_1(r)+ q_1(r)) t(r)=0
$$
with
\begin{eqnarray}
Q_1(r)&=& \frac{h}{16}\frac{r+\beta}{r(r-u_1)(r+1)}+\frac{3}{16}
\left(\frac{1}{(r-u_1)^2}
+\frac{1}{r^2}+\frac{1}{(r+1)^2}\right.\\
&-& \left.2\frac{1}{(u_1+1)(r-u_1)}+2\frac{1}{(u_1+1)(r+1)}\right)
\end{eqnarray}
and
\begin{eqnarray}
q_1(r) &=& \varepsilon \frac{h(1+r)(\beta -1)+6(1+u_1)}{16 u_1(u_1-r)(1+r)^3}\\
&-&\beta_1\frac{h}{16(u_1-r)r(r+1)}.
\end{eqnarray}
Working with the functions $t_k(r)$ we have
$$
\delta t_k(r)=\frac{1}{w_{12}}[ t_1(r)J_{k2}(r)- t_2(r)J_{k1}(r)]
$$
with
$$
J_{lk}(r) = \int_0^r t_l(x)t_k(x) q_1(x) dx
$$
and repeating the procedure explained above we reach the equation
\begin{equation}\label{kfromJ}
2 w_{12} \delta \kappa
= J_{12}(u_1)+\overline{J_{12}(u_1)}
+\frac{\overline{M_{11}(u_1)}}{\overline{M_{12}(u_1)}}
(\kappa^{-2}\overline {J_{11}(u_1)} + \kappa^2 J_{22}(u_1))
\end{equation}
being $M(u_1)$ the unperturbed monodromy matrix at $u_1$.
The $I_{jk}(-1)$ and $J_{jk}(u_1)$ have the structure
$$
I_{jk}(-1)=\varepsilon E_{jk}+\beta_1 B_{jk},~~~~J_{jk}(u_1)=\varepsilon
H_{jk}+\beta_1 L_{jk}~.
$$
Then equating the r.h.s. of eqs.(\ref{kfromI},\ref{kfromJ}) 
we have, given $\varepsilon$, a
system of two equations in the two unknown $\beta_1^R,\beta_1^I$. As
eqs.(\ref{kfromI},\ref{kfromJ}) 
contain both $\varepsilon$ and $\bar\varepsilon$, we see that $\beta$
will not be an holomorphic function of $\varepsilon$ even though we expect
$\beta^R$ and $\beta^I$ to be real analytic functions of the two variables 
$\varepsilon^R$ and $\varepsilon^I$ for small $\varepsilon$. Now the main 
point is
that at the solution of such system of equations the value of the r.h.s. of
eq.(\ref{kfromI}) ( and thus of eq.(\ref{kfromJ})) has to be real and 
positive. 
As we discussed is
detail in Sec.\ref{degreesoffreedom} reality is a general consequence 
of the nature of the
transformation matrices while positivity derives from the fact that for
$\varepsilon=0$ such ratios are positive.

\section{Deformation of the square}\label{squaredeformation}

In this section we shall apply the formalism developed in the previous section
to treat in terms of quadratures the deformed square for any coupling but small
deformation.

In this case we know both the unperturbed value of the accessory parameter
i.e. $\beta=0$ and the explicity form of the unperturbed solutions $y_k$
given is Sec.\ref{square}.

With regard to the perturbation $q$ for the monodromy around $-1$ and $q_1$ for
the monodromy around $1$ we have
\begin{eqnarray}\label{q1q}
q(u) &=& \varepsilon \frac{h(1-u)-12}{16(1+u)(1-u)^3} 
-\beta_1\frac{h}{16u(1-u^2)}\nonumber\\
q_1(r) &=& 
\varepsilon \frac{-h(1+r)+12}{16(1-r)(1+r)^3}- \beta_1\frac{h}{16r(1-r^2)}
\end{eqnarray}
from which we have $q_1(r) = -q(-r)$. On the other hand for the unperturbed $Q$
we have
$$
Q_1(r) = Q(r)
$$
so  that the unperturbed solutions are the same i.e. $y^{(1)}_k(r) = y_k(r)$. 
Taking into account that 
$$
I_{11}(-1) = e^{\frac{i\pi}{2}}I_{11}(1),
~~I_{12}(-1) = -I_{12}(1),~~ I_{22}(-1) =  e^{\frac{3i\pi}{2}}I_{22}(1) 
$$
and
$$
M_{11}(-1)=M_{11}(1)~,~~~~M_{12}(-1) = e^{\frac{i\pi}{2}} M_{12}(1)
$$
we have from (\ref{kfromI}) and (\ref{kfromJ})  $2w_{12}\delta\kappa=0$ i.e. 
$$
0= I_{12}(1)+\bar I_{12}(1)+\frac{\overline{M_{11}(1)}}{\overline{M_{12}(1)}}
(\kappa^{-2}~\overline{I_{11}(1)}+ \kappa^2 I_{22}(1))
$$
which determines the change of the accessory parameter $\beta$ from the value 
$0$ to the value $\beta_1$. It can be written in terms of the unperturbed
solutions (\ref{y1y2hyper}). 
The result $\delta\kappa=0$ 
is not unexpected as the deformations from
the square for symmetry reasons must give the same value of $\kappa$ for
$\varepsilon$ and $-\varepsilon$. 
Substituting the so obtained value $\beta_1$ in eq.(\ref{q1q}) and
computing the ensuing $\delta y_k$ as we did in Sec.\ref{pertonepoint} 
we obtain the $\phi$ for the deformed square.

\section{Addition of weak sources}\label{backgroundgreen}
 
In reference \cite{MV} the technique developed in 
Sec.\ref{pertonepoint}  enabled the computation in terms of quadratures of the
exact Green function on the sphere
in the background generated by three point sources of arbitrary strength. 
Such a result also gave the semiclassical 4-point function with three arbitrary
charges and a small one in terms of quadratures of hypergeometric functions.
Here we shall apply a similar technique to the computation of the
``symmetric'' Green function on the square in the background of one arbitrary
charge or equivalently  the conformal factor generated by an arbitrary charge
and two small charges in symmetric positions. The result extends immediately to
any symmetric distribution of weak charges. This is a different kind of
perturbation on the exact solution eq.(\ref{squarephi}).

Starting from
$$
Q(u) = \frac{1-\lambda^2}{16(u^2-1)}+\frac{3}{16}\frac{(1+u^2)^2}
{u^2(1-u^2)^2}
$$
the addition of a new singularity of strength $\varepsilon$ at $u=t$
introduces the perturbation
$$
q(u)= 
\varepsilon \left[\frac{(1-\lambda^2)\beta}{16u(u^2-1)}+\frac{1}{(u-t)^2}+
\frac{\beta_t}{2(u-t)}+
\frac{\beta_1}{2(u-1)} +\frac{\beta_{-1}}{2(u+1)}\right]. 
$$
The $\beta$'s are related by the Fuchs conditions
$$
\beta_1+\beta_{-1}+\beta_t=0,~~~~2+\beta_1-\beta_{-1}+t\beta_t=0.
$$
the second one needed to leave the source at $z=0$ unchanged.
It will be useful to choose  as the independent parameters $\beta$ and 
$\beta_t$. For the changes in the solutions $y_j(u)$ we have as before
$$
\delta y_j(u) = y_1(u) \frac{I_{j2}(u)}{w_{12}}-
y_2(u) \frac{I_{j1}(u)}{w_{12}}
$$
with
$$
I_{jk}(u) =\int_0^u q(x)y_j(x)y_k(x)dx.
$$
We shall examine first the monodromy at the new singular point $t$. For a tour
around $t$ we have
\begin{equation}\label{deltaIjk}
\delta I_{kj}(u) = \varepsilon \oint_t y_k(x)
y_j(x)\left[\frac{\beta_t}{2(x-t)}+\frac{1}{(x-t)^2}\right]dx=
i\pi\varepsilon\left[\beta_t y_k(x) y_j(x)+ 2 (y_k(x) y_j(x))'\right]_{x=t}
\end{equation}
as the contribution of the remainder of $q(u)$ is zero, being such terms
analytic and thus the monodromy matrix at $t$ becomes
\begin{equation}\label{Mtmatrix}
\begin{pmatrix}
1+\frac{\delta I_{12}(t)}{w_{12}} & 
-\frac{\delta I_{11}(t)}{w_{12}}\\
\frac{\delta I_{22}(t)}{w_{12}} & 
1-\frac{\delta I_{12}(t)}{w_{12}}   
\end{pmatrix}.
\end{equation}
For the computation of
the change in the monodromies at $1$ and $-1$ the procedure is very similar to
the one explained in Sec.\ref{pertonepoint} for the monodromy at $-1$. 

As we have discussed in Sec.\ref{degreesoffreedom} a necessary condition 
for the realization of monodromy at $-1,0,1,t$ is
\begin{equation}\label{threeratios}
\frac{M_{12}(-1)}{\overline {M}_{21}(-1)}=\frac{M_{12}(1)}{\overline
{M}_{21}(1)}=\frac{M_{12}(t)}{\overline {M}_{21}(t)} 
\end{equation}
and we can spend the two complex parameters $\beta,\beta_t$ to solve them.
The explicit form of such equation is given below in 
eqs.(\ref{equalratios1},\ref{equalratios2}).
We perform now the usual transformation $N(n) =(1+\delta K)K M(n)$, where
$n=-1,~1,t$  
with
$$
K=
\begin{pmatrix}
\kappa^{-1}&0\\
0&\kappa
\end{pmatrix},~~~~\delta K=
\begin{pmatrix}
-\delta\kappa&0\\
0&\delta\kappa
\end{pmatrix}
$$
to reduce the three ratios in eq.(\ref{threeratios}) to modulus $1$.

We have now to prove that the ensuing transformation matrices are all
$SU(1,1)$. Due to the choice of the canonical solutions at $0$ the
transformation in $0$ is always
$$
D=
\begin{pmatrix}
i&0\\
0&-i
\end{pmatrix}.
$$ 
In addition we always have
\begin{equation}\label{groupproperty}
N(-1)DN(1) = N(\infty)N(t)^{-1}.
\end{equation}
The transformation $N(\infty)$ is an elliptic $SL(2,C)$ transformation and also
$N(\infty)N^{-1}(t)$ is elliptic, being $N(t)$ infinitesimally near to the
identity. Thus we have
\begin{equation}\label{gammatrace}
{\rm tr}(N(-1)DN(1))=2\cos \gamma = {\rm real}.
\end{equation}
The form of the new matrices $N(-1)$ and $N(1)$ taking into account their
tracelessness is
$$
N(-1)=
\begin{pmatrix}
m & a e^{i\alpha}\\
\bar a & - m
\end{pmatrix},~~~~
N(1)=
\begin{pmatrix}
n & b e^{i\alpha}\\
\bar b & - n
\end{pmatrix},~~~~
$$
so that eq.(\ref{gammatrace}) becomes
$$
2 e^{i\alpha} {\rm Im}(a\bar b) =2\cos\gamma
$$
which implies $e^{i\alpha}=\pm 1$. Due to the continuity of the perturbation 
we have
$e^{i\alpha}=1$.

As a consequence of the general results of Sec.\ref{degreesoffreedom} 
we have that $m$ and
$n$ are pure imaginary, which assures the $SU(1,1)$ nature of $N(-1)$  and
$N(1)$. With regard to $N(t)$ we have at the values $\beta, \beta_t$ 
for which eqs.(\ref{threeratios}) is satisfied
$$
\frac{N_{12}(t)}{\overline{N}_{21}(t)}=\frac{1}{\kappa^2\bar\kappa^2}
\frac{\beta_t y_1^2(t)+ 2 
(y_1^2(t))'}{\bar\beta_t \bar y_2^2(t)+ 2
(\bar y_2^2(t))'} = 1.
$$  
The last expression can also be written as
$$
\frac{\beta_t y_1^2(t)+ 2 
(y_1^2(t))'}{\bar\beta_t \bar y_2^2(t)+ 2
(\bar y_2^2(t))'} =
\frac{(\beta_t y_1(t)y_2(t)+ 2 
[y_1(t)y_2(t)]')^2-4w_{12}^2}
{|\bar\beta_t \bar y_2^2(t)+ 2 (\bar y_2^2(t))'|^2}>0. 
$$
But then being $w_{12}$ real we also have
$$
\beta_t~ y_1(t)y_2(t)+2[y_1(t)y_2(t)]' = {\rm real}
$$
which combined with eq.(\ref{deltaIjk},\ref{Mtmatrix} ) gives 
$N_{22}(t)=\overline{N}_{11}(t)$ i.e. $N(t)\in SU(1,1)$. It follows from 
the group
property (\ref{groupproperty}) that also $N(\infty) \in SU(1,1)$ and thus we
have fulfilled the monodromy at all singularities. 

The explicit form of the two complex equations (\ref{threeratios}) which
determine $\beta$ and $\beta_t$ is
\begin{eqnarray}\label{equalratios1}
I_{12}(-1)+\bar
I_{12}(-1)+\frac{\overline{M}_{11}(-1)}{\overline{M}_{21}(-1)}
\left(\bar\kappa^{-2}\bar I_{11}(-1)+\kappa^2I_{22}(-1)\right)=\\
I_{12}(1)+\bar
I_{12}(1)+\frac{\overline{M}_{11}(1)}{\overline{M}_{21}(1)}
\left(\bar\kappa^{-2}\bar I_{11}(1)+
\kappa^2I_{22}(1)\right)
\nonumber
\end{eqnarray}
\begin{equation}\label{equalratios2}
\bar \kappa^2\kappa^2=\frac{\beta_t y_1^2(t)+ 
2 (y_1^2(t))'}{\bar\beta_t \bar y_2^2(t)+ 2 (\bar y_2^2(t))'} 
\end{equation}
where $\kappa$ is the unperturbed parameter eq.(\ref{kappa4square}) 
solving the problem of the square and $M(\pm1)$ are the unperturbed 
transformation matrices of the $y_k(u)$ of Sec.\ref{square}.

Notice how eq.(\ref{equalratios2}) already fixes the value of 
$\beta_t$ as it
contains only the unperturbed parameter $\kappa$ and 
eq.(\ref{equalratios1}) determines $\beta$.
Finally
$$
I_{12}(1)+\bar
I_{12}(1)+\frac{\overline{M}_{11}(1)}{\overline{M}_{21}(1)}
\left(\bar\kappa^{-2}\bar I_{11}(1)+
\kappa^2I_{22}(1)\right)=w_{12}(\delta\kappa+\delta\bar\kappa) 
$$
determines $\delta\kappa+\bar\delta\kappa$. We recall that $\kappa$ and
$\delta \kappa$ can be chosen real.

We have with $\kappa_1= \kappa(1+\delta\kappa)$
\begin{equation}\label{etominusphihalf}
e^{-\phi/2} =
\frac{\kappa_1^{-2}(\bar y_1+\delta \bar y_1)
(y_1+\delta y_1)-\kappa_1^2(\bar y_2+\delta \bar y_2) (y_2+\delta y_2)}
{\sqrt{2} |\Pi(u)|^2}
\end{equation}
and taking the derivative with respect to $\varepsilon$ we have
\begin{eqnarray}\label{greenfunc}
& &-\frac{1}{8\pi}
\frac{\partial \phi}{\partial \varepsilon}= G(z,t))=\nonumber\\
& &\frac{1}{4\pi(\kappa^{-2}\bar y_1 y_1-\kappa^2 y_2 y_2)}\left[
\left(-2\kappa' + \frac{{\cal I}_{12}}{w_{12}}+
\frac{\overline{\cal I}_{12}}{w_{12}}\right)
(\kappa^{-2}\bar y_1y_1+\kappa^2\bar y_2y_2)-\right.\\
& &\left.\bar y_2 y_1\left(\kappa^{-2}\frac{\overline{\cal I}_{11}}{w_{12}}+
\kappa^2 \frac{{\cal I}_{22}}
{w_{12}}\right)-
y_2\bar y_1 \left(\kappa^{-2} \frac{{\cal I}_{11}}{w_{12}}+
\kappa^2 \frac{\overline{\cal I}_{22}}{w_{12}}\right)\right]\nonumber
\end{eqnarray}
where
$$
{\cal I}_{jk} =\int_0^u \frac{q(x)}{\varepsilon} y_j(x) y_k(x)dx~~~~{\rm
and}~~~~\kappa'=\frac{\delta \kappa}{\varepsilon}. 
$$
This is the exact symmetric Green function in the background generated by the 
one point
source of arbitrary strength on the square, given in terms of quadratures.
It satisfies the equation
\begin{equation}\label{greenequation}
(\Delta -e^{\phi})G(z,t) = \delta^2(z-z_t)+\delta^2(z+z_t)
\end{equation}
being $\phi$ the background generated by the one point source placed at
$z=0$ and $z_t$ is the image on the $z-$plane of $t$. It is possible
to verify explicitely eq.(\ref{greenequation}) by taking the derivatives 
of  expression (\ref{greenfunc}) and using the relations
$$
I_{jk}'(u) = q(u) y_j(u) y_k(u).
$$
$G$ has a singularity in $z=\pm z_t$ but it is regular at $z=0$ which is the
position of the source as in happens in the sphere topology \cite{MV}.
The quantity $e^{\phi}$ with $\phi$ given by eq.(\ref{etominusphihalf}) is 
the semiclassical two point function with a source of finite strength and 
the other two of small strength.

\section{Conclusions}\label{conclusions}

In this paper we have applied in a systematic way the Riemann-Hilbert technique
to Liouville theory on the torus. After a change of coordinates the problem is
reduced to a Fuchsian differential equation. In the case of the one point
function such differential equation is a particular case of the Heun
differential equation, i.e. a differential equation with four singularities,
which as well known, contains an accessory parameter which is not fixed by the
Fuchs conditions. Then the main problem is to determine the value of such
parameter. This is performed here in a few cases, through a perturbative
technique, while in the case of the square the problem can be solved exactly.

The transformation through the Weierstrass function $\wp(z)=u$ gives a double
valued representation of the torus in the $u$-plane. In the present paper we
have examined situations which are symmetrical, where is not necessary to
distinguish among the two sheets. We shall examine in an other paper the
asymmetric situation.

The case of the one point function on the square is explicitely solvable
through hypergeometric functions.  The perturbative cases treated in the
present paper are given by one weak source, by a general deformation and by
the addition of a weak symmetric distribution of charges.  Formulas for a
general deformation are derived and then applied explicitely to the
deformation of the square. An other application of the perturbative technique
is the determination, through quadratures, of the exact symmetric Green
function on the background generated by the one point source of arbitrary
strength. The technique developed here can be applied to the semiclassical
expansion of quantum Liouville theory on the torus.

\section*{Acknowledgments}

I am grateful to Massimo Porrati for arising my interest in the problem and
for useful discussions.

\vfill

%101024.1

\end{document}